# Dynamical sensitivity control of a single-spin quantum sensor

Andrii Lazariev[1], Silvia Arroyo-Camejo[2], Ganesh Rahane[1], Vinaya Kumar Kavatamane[1] and Gopalakrishnan Balasubramanian[1*]

[1] Max-Planck Research Group Nanoscale Spin Imaging, Max Planck Institute for Biophysical Chemistry, Göttingen, GERMANY.

[2] Department of Nanobiophotonics, Max Planck Institute for Biophysical Chemistry, Göttingen, GERMANY.

**The Nitrogen-Vacancy (NV) defect in diamond is a unique quantum system that offers precision sensing of nanoscale physical quantities**[1–8] **beyond the current state-of-the-art. Here we present a method to controllably encode the interactions in the population of the spin states, thereby introducing a way to control the sensitivity of a single spin as a continuum in contrast to free-evolution based methods. By adopting this feature we demonstrate high-accuracy NV magnetometry without $|2\pi|$ ambiguities, enhance the dynamic range by a factor of $4 \cdot 10^3$, achieve interaction times exceeding 2 ms in off-the-shelf diamond. We perform nuclear spin-noise spectroscopy in the frequency domain by dynamically controlling the NV spin's sensitivity piecewise and in a smooth manner thereby precluding harmonic artefacts and undesired interactions**[9,10]. **On a broader perspective dynamical sensitivity control provides an elegant handle on the inherent dynamics of quantum systems**[11], **while offering decisive advantages for NV centre applications notably in quantum controls**[12,13] **and single molecule NMR/MRI**[14–16].

* Email: G. Balasubramanian (gbalasu@mpibpc.mpg.de)

Despite the cutting edge advances using NV sensors for probing nanoscale physical quantities a few factors still repress widespread adaptations. Some of those impediments are the required foreknowledge/control on the quantity to be measured, ruinous effects of environmental noise, limited dynamic range (DR) and ambiguous spectral information. Here we demonstrate dynamical sensitivity control (DYSCO) method and applications in NV metrology that mitigate some of the hurdles mentioned above. The DYSCO pulse scheme encodes a magnetic field dependent shift on the spin state population, the magnitude of change can be varied by a control parameter of the field that drives the spin. This provides us means to modulate the sensitivity of a single spin to the external field by circumventing free-precession[9]. In optical terminology, this method depicts a polarimetric measurements scheme and complements interferometric methods that are routinely employed for sensing. In the context of metrology[17], polarimetry schemes are generally considered to be robust against drifts and fluctuations. As an added feature we show that the piecewise dynamical control of the spin allows for frequency selective sensing in a lock-in manner without compromising the inherent robustness of polarimetric schemes.

As spin probe, we employ the NV centre ground state triplet system featuring the spin states $|-\rangle$, $|0\rangle$, and $|+\rangle$ (details in Supplementary Information). The NV spin is coherently driven on the $|0\rangle$ to $|-\rangle$ transition by a microwave field (MW) of amplitude $B_1$, detuning $\delta_-$ and phase $\varphi$. In the presence of an external weak RF magnetic field $B_{\text{RF}}(t)$ the system interaction Hamiltonian (after the rotating wave approximation) has the form

$$H(t) = -\delta_- |-\rangle\langle-| - \gamma_{\text{NV}} B_{\text{RF}}(t) \cdot (|+\rangle\langle+| - |-\rangle\langle-|) - \frac{\Omega_-}{2} \left( e^{i\varphi(t)} |0\rangle\langle-| + \text{h.c.} \right),$$

where $\gamma_{\text{NV}} \approx -2\pi \cdot 28$ GHz/T is the NV gyromagnetic ratio and $\Omega_- = -\gamma_{\text{NV}} B_1$ is the Rabi frequency of the driving microwave field with the phase $\varphi$.

The dynamical sensitivity control scheme is composed of $N$ control units each consisting of $4 \cdot \pi$-pulses that sequentially drive the NV spin state in a phase alternating manner:

$$[\pi_{\bar{x}-\varphi}, \pi_{x-\varphi}, \pi_{x+\varphi}, \pi_{\bar{x}+\varphi}]^N - \pi_y - [\pi_{x+\varphi}, \pi_{\bar{x}+\varphi} \pi_{\bar{x}-\varphi}, \pi_{x-\varphi}]^N.$$

A $\pi_y$-pulse placed in the middle of the sequence followed by $N$ reversely ordered $4 \cdot \pi$-pulses compensates for pulse-errors (cf. Figure 1a). The total evolution time is given by $t_N = (4N + 1/2) \cdot 2\pi/\Omega_-$ and its reciprocal defines the spectral resolution. The $4 \cdot \pi$-pulse units are phased and arranged such that the spin vector returns to the same position on the Bloch sphere when it is not interacting with additional fields (green trajectory in Figure 1a, Simulations). If, however, an external $B_{RF}$ field is present, it will render the spin trajectory in such a way that initial and final spin states are not identical anymore. The net effect of the $B_1$ and $B_{RF}$ fields imparts a population shift on the final spin state (individual $\pi$ pulses are shown in red, blue, magenta, cyan trajectories in Figure 1a, Simulations), which is dependent on the magnitude of the external field $B_{RF}$ and the phase angle $\varphi$ of the $4 \cdot \pi$-pulse unit. To visualize the dynamics of the state evolution resulting from one $4 \cdot \pi$-pulse unit we reconstructed the spin vector at few instances of time from $x, y$ and $z$ projections of the spin on the Bloch sphere spanned by the $|0\rangle$ and $|-\rangle$ states. We show the trajectories of the spin vector associated with the $4 \cdot \pi$-pulses (shown in red, blue, magenta, cyan) for a phase angle of $\varphi = \pi/6$ and observe that the $B_{RF}$ field influences the evolution causing a net change in population (cf. Figure 1a, Experiment).

To illustrate this significance in the context of NV magnetometry we simulate the position of the state vector at the end of the sequence to visualize its dependence on the $B_{RF}$ field strength (cf. Figure 1b). While a Hahn-echo sequence[3,18] accumulates the $B_{RF}$ influence on the phase of the superposition state $\frac{1}{\sqrt{2}}(|0\rangle + |-\rangle)$, the DYSCO method encodes the magnetic field interactions on the change of the population of the states $|0\rangle$ and $|-\rangle$. In Figure 1c we plot the occupancy of the $|0\rangle$ spin state $P_0$ measured after the completion of the pulse sequence (with $N = 1$) and its dependence on the strength of the $B_{RF}$ field and the phase angle $\varphi$ of the driving pulses as derived from the interaction Hamiltonian. These theoretical expectations from Figure 1c agree very well with our experimental results shown in Figure 1d, in which we concatenated $N = 20$ of the $4 \cdot \pi$-pulse units and ramp the $B_{\mathrm{RF}}$ field amplitude to obtain a few oscillations.

The sensitivity of a measurement is given by the smallest change in the quantity to be measured (here $\delta B_{\mathrm{RF}}$) that still leads to a resolvable change of the experimental observable[1] (here $\delta P_0$). Both in the case of conventional and DYSCO magnetometry the population $P_0(B_{\mathrm{RF}})$ varies harmonically with increasing magnetic field amplitude $B_{\mathrm{RF}}$, and the sensitivity can be obtained from the maximum slope of the response[3] $dP_0/dB_{\mathrm{RF}}$. As the signal is harmonic, the Fourier transform of the population with respect to $B_{\mathrm{RF}}$ given by $S(\zeta) \propto \mathcal{F}(P_0(B_{\mathrm{RF}}))$ features a single Dirac delta distributed $\zeta$-component which is a measure for the sensitivity. We investigate the DYSCO method in the context of magnetometry, using $N = 200$ of the $4 \cdot \pi$-pulse units and study the dependence of $P_0(\varphi, B_{\mathrm{RF}})$ for $\varphi \in [0, \pi]$. We observe that $P_0(\varphi, B_{\mathrm{RF}})$ responds harmonically in the range $(0 < \gamma B_{\mathrm{RF}} \leq \Omega/4)$, and present the corresponding Fourier transform $S(\zeta)$ for every value of $\varphi \in [0, \pi]$ in Figure 2a (top). Thus, the sensitivity dependence of the DYSCO sequence can be deduced from the experimental data using $\beta(\varphi) \propto \left|\mathrm{argmax}_{\zeta}\left[\mathcal{F}(P_0(B_{\mathrm{RF}}))\big|_\varphi\right]\right|$, we denote $\beta(\varphi)$ as the dynamical sensitivity in the following, because this quantity can be continuously varied as desired by means of the phase angle $\varphi$ of the $4 \cdot \pi$-pulse units. Note, that here a value $\beta = 0$ stands for minimal sensitivity, while a value of $\beta = 1$ represents maximal sensitivity. Empirically we find that the dynamical sensitivity has a dependence on the MW phase described by $\beta(\varphi) \propto |\sin \varphi|$. Complementary, we used *SpinDynamica*[19] to simulate the response of the NV spin on the $B_{\mathrm{RF}}$ field and the phase angle of the pulse sequence for $N = 160$ of the $4 \cdot \pi$-pulse units (for details see Supplementary Information). These simulations are shown in Figure 2a (bottom) and correspond well with the experimentally obtained results.

In the two-dimensional subspace spanned by $|0\rangle$ and $|-\rangle$ the spin vector dynamics can be represented by rotation matrices. Using successive rotations corresponding to the DYSCO pulse sequence we derive an explicit expression for $P_0(\varphi, B_{RF})$. From the series expansion of $P_0(\varphi, B_{RF})$ we obtain an analytical expression for the dynamical sensitivity as $\beta(\varphi) \propto \left(\frac{B_{RF}}{B_1}\right)\sin\varphi$. This is consistent with the experimental results (details in Supplementary Information). We observe that the resultant population shifting effect of the DYSCO sequence is almost identical when the spin state is initially prepared in a superposition state (see Supplementary Information). For an intuitive understanding an illustrative representation of the spin dynamics in terms of rotations on the Bloch sphere is detailed in the Supplementary Information.

As an application, we show that the dynamical sensitivity control method can be used to enhance the dynamic range of an NV magnetometer. The DR is the extent of the magnetic field that can be measured without $|2\pi|$ ambiguities arising from the oscillatory nature of the $P_0(B_{RF})$ response[20–22]. The control method presented here gives us the ability to gradually ramp the dynamical sensitivity of the spin $\beta_k$ between 0 and 1 in arbitrary steps and to measure the magnetic field $B_{RF}$ influence on the NV spin free from $|2\pi|$ ambiguity as shown in Figure 2b. Parameters that bound the dynamical control of the NV spin are the Rabi frequency $\Omega_- = \gamma_e B_1$ and the spin relaxation time in the rotating frame $T_{1\rho}$ that limits the interrogation time of DYSCO and denoted by $T_{\text{DYSCO}}$ (Figure 2c). For a single spin magnetometer we achieve a DR of about $4 \cdot 10^3$, obtained from the ratio of the respective maximum slopes of the $P_0(B_{RF})|_{\varphi \to 0}$ and the $P_0(B_{RF})|_{\varphi \to \pi/2}$ responses shown in Figure 2d. The permissible bounds of the DR are given by the ratio of the largest and smallest fields $B_{RF}$ field changes that can be measured by the sensor. The lower bound is obtained by setting $\beta = 1$, whereas $\beta = 0$ provides an upper bound for the DR. For our experimental conditions with $\Omega_- = 2\pi \cdot 8.33$ MHz and $T_{\text{DYSCO}} = 2.55$ ms a theoretical maximum DR of $5 \cdot 10^3$ is achievable. Furthermore, the frequency selectivity of the dynamical control allows sensing without being hindered by $^{13}$C spin-bath signatures[23] (cf. Figure 2c). The DYSCO method permits the NV sensor to operate over a wide frequency bandwidths ranging from $\Omega_-/9\pi$ to $1/T_{\text{DYSCO}}$ set by the minimum $(4N + 1/2) \cdot 2\pi/\Omega_-$ and $T_{\text{DYSCO}}$ maximum permissible duration $t_N$ of the sensing scheme. We show that by using electronic grade diamond (1.1% $^{13}$C) the DYSCO scheme provides an elegant route to achieve longer interrogation times to a maximum of $T_{\text{DYSCO}} = 2.55\text{ms}$ that approaches $T_{1\rho} = 3.2\text{ ms}$ and a corresponding enhancement in the sensitivity that was previously considered exclusive to $^{12}$C isotope enriched diamond[18] (cf. Figure 2c,d).

We show the piecewise dynamical control of the sensitivity of a spin sensor allows us to measure strength and frequency components of arbitrary oscillatory fields (details in Methods section). In Figure 3a, we present the experimental results of the NV sensor response to an external field oscillating at 8 kHz. As the dynamical sensitivity of the sensor is modulated smoothly, the scheme precludes any harmonic response as those arising in interferometric multi-pulse schemes[9,10] (cf. Figure 3b and methods section for filter function representation). The method is useful for resolving frequency components of multiplexed signals, because the sensitivity can be tuned in a controlled manner to detect signals with narrow bandwidth. In the results shown in Figure 3c, we inject a set of six frequencies (2, 4, 5, 6, 7, and 9 kHz) of various strengths and measure the response in a spectrogram fashion with a resolution of 1 kHz.

Probing phase asynchronous signals is particularly interesting in the context of sensing nuclear spin noise[24]. We realized this asynchronous measurement scheme as before by dynamically modulating the sensitivity $\beta(t_n)$ with a desired frequency $f_s$ and gradually varying the maximum of the dynamical sensitivity $\beta_k$ in steps (details in methods section). A phase asynchronous signal with a frequency of 100 kHz is injected into a micro-coil, and the spectral response of the NV spin $P_0$ is plotted in Figure 3d as colour coded spectrogram. At the maximum response, the spin state attains a value of $P_0 = 1/2$ indicating incoherent interaction. The results also exhibit the absence of harmonic responses (cf. Figure 3d).

In another set of experiments, we demonstrate the use of the DYSCO method to probe noise spectral details of the $^{13}$C spin bath in the vicinity of a single NV spin. The $^{13}$C nuclear spins in the bath process with a certain spread in Larmor frequencies. This causes asynchronous magnetic field fluctuations influencing the NV spin. As detailed above we modulate the NV spin's dynamical sensitivity in an extended range from 10 kHz to 1000 kHz with 10 kHz resolution and record the response (cf. Figure 4a). It is evident from Figure 4a that the $^{13}$C nuclear spin noise signatures are seen only at the $^{13}$C Larmor frequency (430 kHz), without any harmonic or spurious artefacts[10]. To emphasize the absence of harmonics, we injected an additional 100 kHz RF noise signal and observed the response at the expected frequencies. Moreover, the injected noise at 100 kHz did not influence the $^{13}$C Larmor response occurring roughly at the 4th harmonic of the injected signal (Figure 4b). The frequency resolved method using DYSCO thus permit sensing $^1$H nuclear spins without interference from $^{13}$C nuclear spins that precess at 1/4 the Larmor frequency of $^1$H spins[10] ($\gamma_{1H}/\gamma_{13C} \sim 4$). Applications of spin based molecular scale NMR/MRI[14–16] in which NV resides close to surfaces yet requires high sensitivity and selectivity would excessively benefit from employing the DYSCO sensing scheme. A high-resolution scan in the frequency range from 380 kHz to 500 kHz near the $^{13}$C Larmor frequency shows well-resolved signatures of the NV spin coupled to distant $^{13}$C nuclear spins with a coupling strength as small as 30 kHz (Figure 4c,d). The frequency domain sensing scheme that delivers high-resolution noise spectra will be of profound use in numerous solid-state quantum architectures to identify spin resources, possible dissipation pathways and protecting the qubits[25,26]. The piece-wise dynamical control provides an alternate scheme to protect the central-spin and realize high-fidelity quantum operations[12] similar to optimal-control protocols[13].

In summary, the DYSCO method provides a framework for preserving the sensitivity of a quantum system despite a noisy environment, while allowing for the modulation of its effective coupling to desired external field in a controlled and robust fashion. Our method to boost the dynamic range and to allow sensing of weak signals without preconditions will benefit real-life applications of NV sensors[27], while it can be also adopted for other quantum systems[11,25,26]. The DYSCO route to achieve $T_{1\rho}$ limited sensing times and to detect sub-kHz coherent interactions without resorting to spin-free isotopic enrichment provides an accessible NV quantum technologies based on affordable materials[18]. While the results demonstrate decisive advantages of our method in NV spin magnetometry, it is seamlessly applicable for the measurement of other relevant physical quantities and even in other NV sensing modalities[6]. The DYSCO method presented here provides a vital step towards realizing the full potential and uniqueness of the NV sensor as a tool for examining nanoscale phenomena and processes, addressing the challenges and technological demands of the future.

**Methods**

**Sample and setup:** For the experiments we used a single NV centre in an electronic grade CVD grown diamond (Element Six). The spin manipulation is done in a home-built confocal microscope equipped with an arbitrary waveform generator synthesizing microwave fields at 20 GS/s. In our diamond sample, which contains natural abundance of 1.1% $^{13}$C nuclear spins, we selected an NV centre that did not show signatures of strongly coupled $^{13}$C nuclear spins. A magnetic bias field of $B_0 = 40.4$ mT is aligned parallel to the NV centre axis to polarize the $^{14}$N spins, resulting in an Optically Detected Magnetic Resonance (ODMR) line width of about 100 kHz. The NV spin is driven with a Rabi frequency of 8.33 MHz using a nearfield antenna, and an RF micro coil placed in the vicinity provides the desired $B_{\text{RF}}$ magnetic field.

**Sensing magnetic signals using dynamical sensitivity control:** In this spectroscopy approach the desired spectral resolution specifies the required signal acquisition time $t_N$. A number of $N$ times $4 \cdot \pi$-pulse units concatenated for the total duration $t_N$. Each of these $4 \cdot \pi$-pulse units can be composed to modulate the instantaneous dynamical sensitivity of the NV spin sensor to match any arbitrary temporal profile of the frequency $f_{\text{RF}}$ to be sensed. The modulation is discretized by the $4 \cdot \pi$-pulse unit length, thus higher Rabi frequency can provide smoother modulation. These piece-wise sensitivity control using these discretized units enables frequency resolved sensing (spectroscopy) of arbitrary oscillating magnetic fields. For a measurement of the spectral components of a signal in the range of frequencies $f_s \in [f_{\min}, f_{\max}]$ with a resolution defined by $t_N$, the total sensing time ($t_N$) is kept fixed and sequentially for each $f_s$ the DYSCO modulation pattern is composed to be symmetrical around the middle $\pi_y$-pulse. The modulation is composed such that the dynamical sensitivity of the $n^{\text{th}}$ individual $4 \cdot \pi$-pulse unit is defined as $\beta(f_s, t_n) = \beta_k \sin(2\pi f_s t_n)$ with $t_n = (1 + 2n) \cdot 2\pi/\Omega_-$ and $n \in \{0,1,2 \dots, N\}$. The amplitude factor $\beta_k \in [0,1]$ is constant for every $k$-th experimental run and can be stepwise increased with $k \in \{0,1,\ldots,K\}$ which removes of the typical $|2\pi|$ ambiguity of interferometric sensing methods. Complete modulation periods are symmetrized by including $\pi_y$-pulses, and if incommensurate portions are present they appear symmetrically. In the spectral domain, if the $\sin x/x$ wiggles are caused due to a rectangular window $\beta_k = k/K$, then by using a Gaussian envelop $\beta_k(t_n) = k/K \cdot \exp\left(-\frac{(t_n - t_N/2)^2}{2(t_N/2)^2}\right)$ we can supress these wiggles. The DYSCO magnetometry experiments are performed by measuring the NV signal $P_0$ and varying the modulation frequency of the dynamical sensitivity $\beta(f_s)$ along the abscissa and the maximum value of dynamical sensitivity $\beta_k$ along the ordinate. The sensor responds when the modulation frequency of the dynamic control $f_s$ matches the frequency $f_{\mathrm{RF}}$ of the external $B_{\mathrm{RF}}$ field. At that modulation frequency $f_s$ upon variation of the maximum sensitivity $\beta_k$ (the ordinate) the resulting $\mathcal{F}(P_0(\beta_k))$ response gives a measure of the magnetic field strength (cf. Figure 3a,c). If the signals are phase synchronized, the spin state oscillates with a rate proportional to the strength of the magnetic field $B_{\mathrm{RF}}$. As the dynamic sensitivity of the sensor is ramped from zero every oscillation is observed and the measurement is free of $|2\pi|$ ambiguity. If the external field happened to be asynchronous and when $\beta_k$ is varied the signal gradually drops from $P_0 = 1$ and reaches a value of half the visibility $P_0 = 1/2$. Another variant of the sequence offers greater flexibility in a wide frequency span. The sequence is constructed with one $\pi_y$ pulse in the middle of the evolution time and a number of $N$ times the $4 \cdot \pi$-pulse modulation units. The DYSCO noise spectrum is acquired as explained before in the frequency domain by measuring the signal $P_0$ and probing a range of modulation frequencies and varying the maximum of dynamical sensitivity $\beta_k$. It should be noted that a single frequency scan with a fixed value of the sensitivity maximum $\beta_k$ can also show the presence of spectral components, but without details on their magnitude.

**Filter functions and harmonics free sensing:** The frequency and selectivity of a spin sensing protocol are governed by a filter function $F(\omega)$ characteristic of a pulse sequence. The $F(\omega)$ function is deduced from a sensitivity function $g(t)$ that describes the instantaneous sensitivity of the spin during the evolution time[9]:

$F(\omega) = \frac{\omega^2}{2}|\mathcal{F}(g(t))|^2$. The NV spin coherence signal $\chi(t)$ is influenced by the spin noise spectrum $S(\omega)$ in the environment and given by

$$\chi(t) = \int_0^\infty \frac{d\omega}{\pi} S(\omega) \frac{F(\omega t)}{\omega^2}$$

The filter functions $F(\omega)$ of the sensing protocols that accumulate phase by free evolution and multiple $\pi$-pulses suffer from undesired harmonic responses[10] as shown in Figure 3b. Harmonics free sensing resulting from the DYSCO protocol can be explained in a spectroscopic representation by using filter functions. As we modulate the dynamical sensitivity $\beta(t)$ in a smooth way, the Fourier transform gives a single-valued frequency response (shown in Figure 3b). Either by sequentially scanning the desired frequency range or by non-uniform sampling we can perform weak precision magnetic sensing in the frequency domain.

**Methods to mitigate harmonic artefacts:** The DYSCO method presented here utilizes dynamic modulation of the driven NV spin. Some methods that use dynamical decoupling together with cyclic phases and non-uniform free-precession intervals are proposed for suppressing harmonic responses[28]. These methods could incorporate the DYSCO modulation mechanism to prevent abrupt jumps in the sensitivity function and prolong the high frequency limits. Schemes that employ active manipulation of nuclear spins[29], and spin-lock[30] can also mitigate harmonics while being applicable in limited bandwidth naturally precluding multiplexing capabilities.

26. Bylander, J. *et al.* Noise spectroscopy through dynamical decoupling with a superconducting flux qubit. *Nat Phys* **7,** 565–570 (2011).

27. Le Sage, D. *et al.* Optical magnetic imaging of living cells. *Nature* **496,** 486–489 (2013).

28. Casanova, J., Wang, Z.-Y., Haase, J. F. & Plenio, M. B. Robust dynamical decoupling sequences for individual-nuclear-spin addressing. *Phys. Rev. A* **92,** 042304 (2015).

29. Mamin, H. J. *et al.* Nanoscale nuclear magnetic resonance with a nitrogen-vacancy spin sensor. *Science* **339,** 557–60 (2013).

30. Loretz, M., Rosskopf, T. & Degen, C. L. Radio-Frequency Magnetometry Using a Single Electron Spin. *Phys. Rev. Lett.* **110,** 017602 (2013).
**Acknowledgements** We thank Prof. S.W. Hell for discussions and support in the project. We thank Prof. P.K. Madhu, Prof. M. Levitt and Prof. C. Griessinger for the insights and contributions on developing the sequences. We thank Prof. E. Sjöqvist and Dr. D. Mesterhazy for discussions on the mechanism of DYSCO. We thank Prof. F. Jelezko and Dr. J. Mamin for discussions on applications of DYSCO method. We gratefully acknowledge funding from the Max-Planck Society, Niedersächsisches Ministerium für Wissenschaft und Kultur and DFG Research Centre Nanoscale Microscopy and Molecular Physiology of the Brain and the DARPA QuASAR program.

**Author Contributions** All the authors contributed towards all aspects of the project and the article. G. B. supervised the project and wrote the manuscript.

**Competing Financial Interests** The authors declare no competing financial interests.

**Correspondence and requests for materials** should be addressed to G. B. (gbalasu@mpibpc.mpg.de).

**Supplementary Information** accompanies this version of the paper.
13

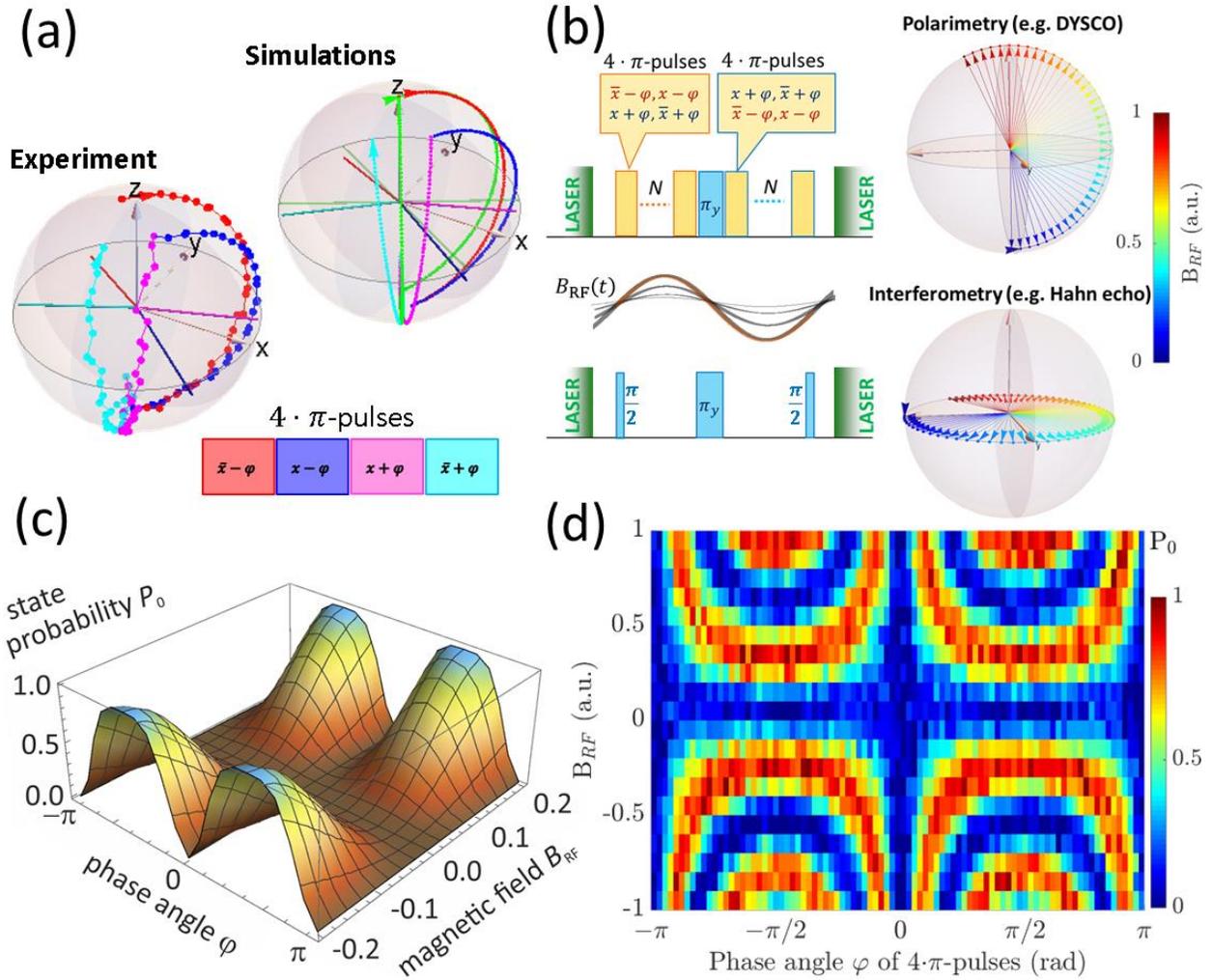

**Figure 1 | Dynamical sensitivity control (DYSCO).** a) Schematic representation of a DYSCO $4 \cdot \pi$-pulse unit. Simulations and experimentally traced spin vector trajectory for one $4 \cdot \pi$-pulse unit represented on the Bloch sphere spanned by $|0\rangle$ and $|-\rangle$. The green trajectory in simulations depicts spin evolution when sequentially driven by $4 \cdot \pi$-pulse and $B_{RF} = 0$. The coloured trajectories in the Simulations and Experiment are shown for $B_{RF} \neq 0$. Individual $\pi$-pulses are shown in red, blue, magenta and cyan dots and traces. b) Schematic of the DYSCO method and the expected changes in the final state population $P_0$ for different values of the $B_{RF}$ field amplitude shown in comparison to the Hahn-echo sensing method (note the last $\pi/2$ pulse used in NV-metrology based on free-precession/interferometry is not included here for vividly contrasting the sensing mechanism of DYSCO). c) Explicit calculation of the level occupancy $P_0$ and its dependence on the $B_{RF}$ field and the phase angle $\varphi$ of the driving pulses at the end of the pulse sequence for $N=1$ (the $B_{RF}$ field is given in units of $\Omega_-$). d) Experimental results showing the dependence of state population $P_0$ as a function of $B_{RF}$ and $\varphi$ for $N = 20$.

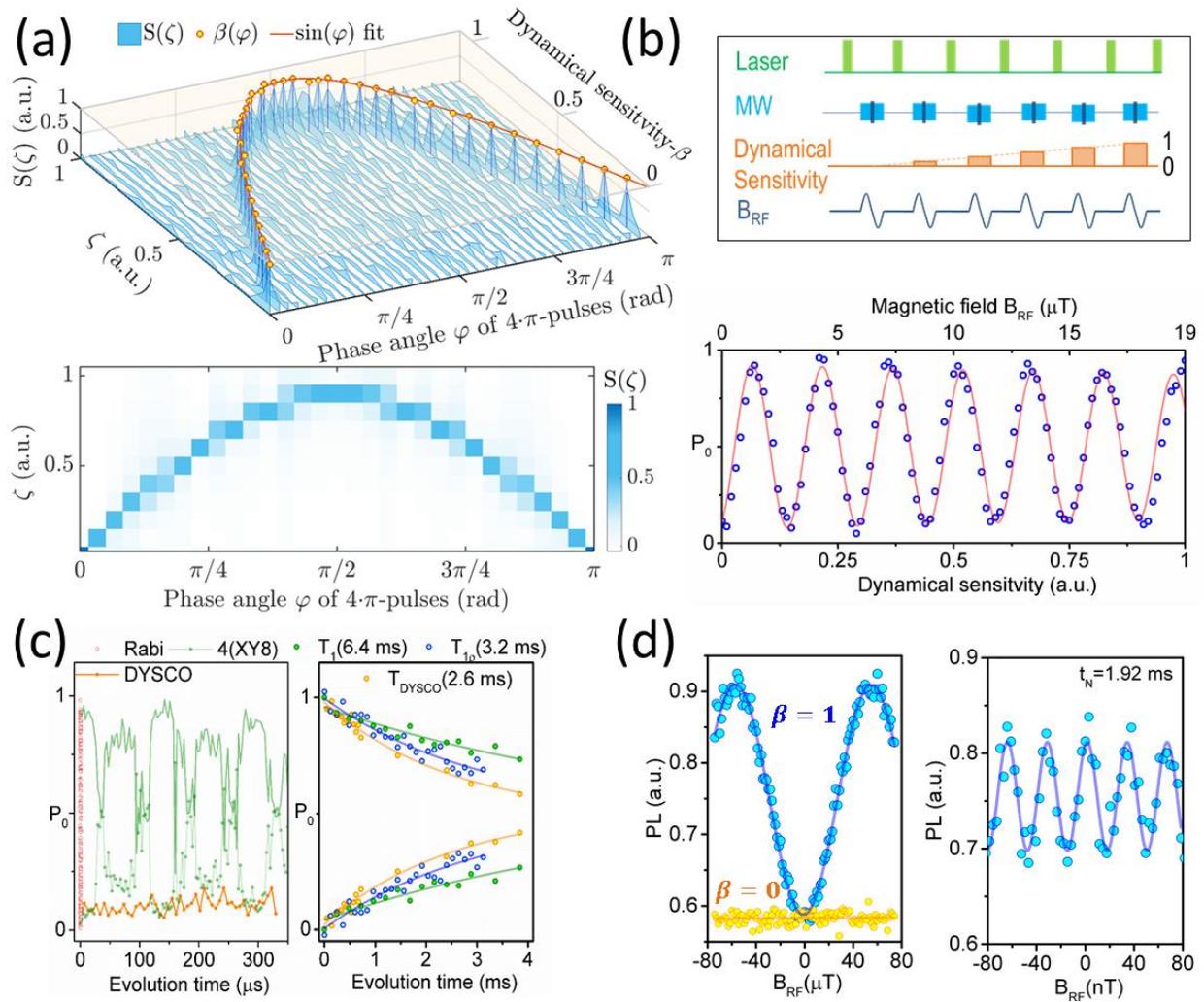

**Figure 2 | Results of dynamical sensitivity and dynamic range enhancement.** a) Experimental results showing the $S(\zeta)$, dynamical sensitivity $\beta(\varphi)$ and its dependence on the $4 \cdot \pi$-pulse unit phase angle $\varphi$ (upper plot). Simulations showing the $S(\zeta)$ dependence on $\varphi$ obtained close to experimental conditions (lower plot). b) Schematic and the results of magnetic sensing without $|2\pi|$ ambiguity by controlling the dynamical sensitivity. c) Experimental results of an XY8-4 dynamical decoupling together with DYSCO clearly showing absence of characteristic $^{13}$C nuclear spin influence (left plot). Experimental data and fits showing the relaxation times $T_1$, $T_{1\rho}$ and $T_{\text{DYSCO}}$ of a single NV spin (right plot). d) Results depicting the extent of the dynamic range ($4 \cdot 10^3$) achieved using DYSCO (left plot). Precision magnetometry performed at $t_N = 1.92$ ms total acquisition time using an NV spin present in a natural abundancy 1.1% $^{13}$C diamond (right plot).

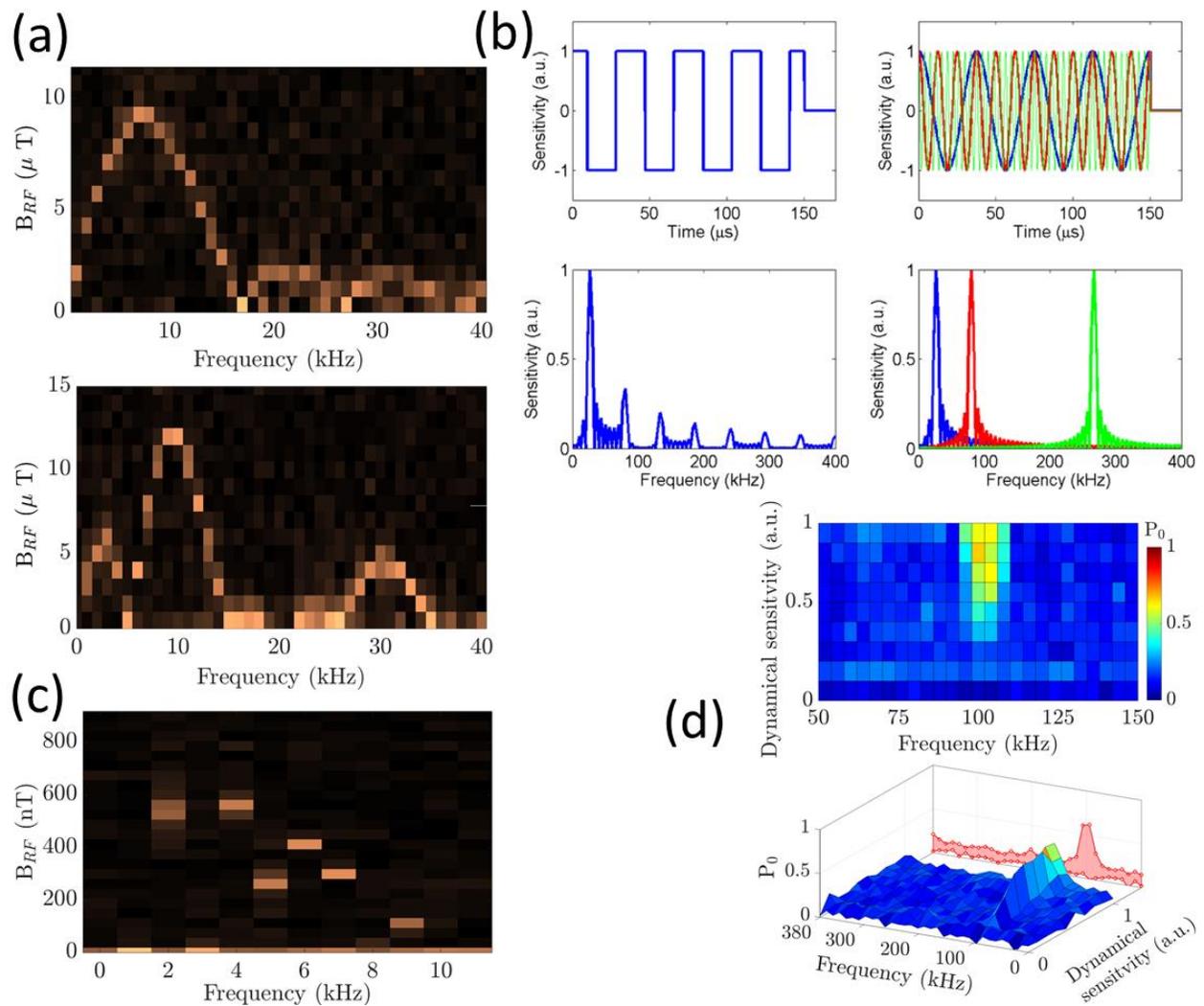

**Figure 3 | High-resolution magnetic sensing in the frequency domain.** a) Frequency domain sensing of an 8 kHz phase-synchronized signal using DYSCO and conventional sensing by multi-pulse scheme. b) Free precession based multi-pulse sequences (upper left) produce harmonics in their filter function (lower left), while the DYSCO sequence (upper right) produces a single frequency response (lower right). c) Experimental DYSCO spectrum of six-multiplexed signals with 1 kHz spectral resolution. d) Frequency domain response of an asynchronous 100 kHz noise signal.

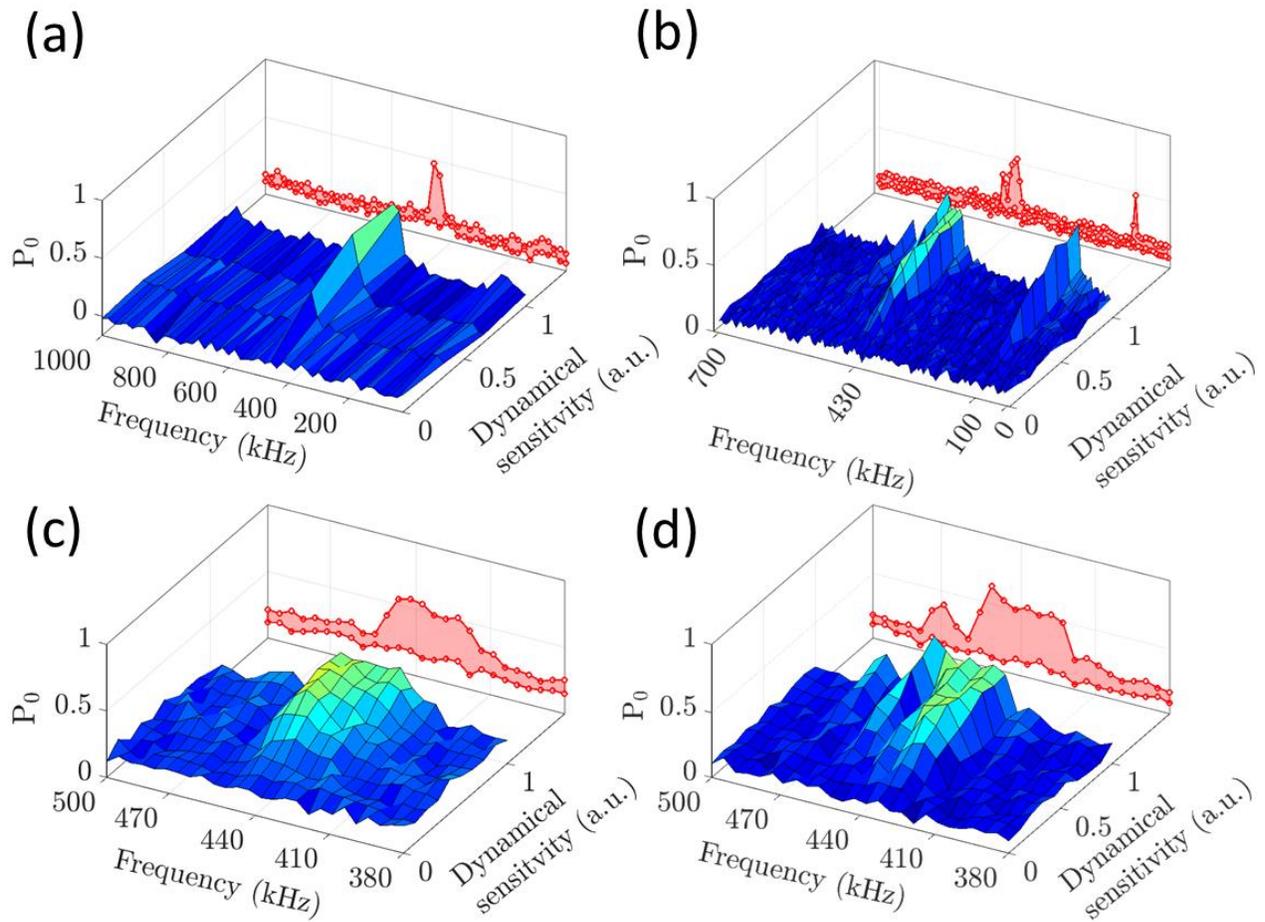

**Figure 4 | High-resolution noise spectrum of $^{13}$C nuclear spin bath in diamond free from harmonics.**
a) The spin noise spectrum of the $^{13}$C nuclear spin bath, shows distinctly the carbon Larmor frequency in a wide range without any harmonics. b) The $^{13}$C noise spectrum together with 100 kHz noise signal showing no undesired responses. Finer resolved frequency scans with a resolution of c) 10 kHz and d) 5 kHz showing signatures of $^{13}$C spins being weakly coupled to the NV spin with a strength of 30 kHz.